\documentstyle[psfig]{mn2e}
\newif\ifAMStwofonts

\newcommand{\K}{K$^{\prime}$}

\title[]
{The relative abundances of ellipticals and starbursts 
among the Extremely Red Galaxies.
}

\author[Mannucci et al.]
{
F. Mannucci$^1$, 
L. Pozzetti$^2$, 
D. Thompson$^3$,
E. Oliva$^{4,5}$,
C. Baffa$^4$,
G. Comoretto$^4$,
\newauthor
S. Gennari$^4$, and
F. Lisi$^4$\\
$^1$ C.A.I.S.M.I. - C.N.R., Largo E. Fermi 5, I-50125, Firenze, Italy \\
$^2$ Osservatorio Astronomico di Bologna, via Ranzani 1, I-40127,
Bologna, Italy\\
$^3$ Palomar Observatory, California Institute of Technology, MS 320-47,
Pasadena, CA 91125\\
$^4$ Osservatorio Astrofisico di Arcetri, Largo E. Fermi 5, I-50125,
Firenze, Italy \\
$^5$ Centro Galileo Galilei \& Telescopio Nazionale Galileo, P.O. Box
565, E-38700, Santa Cruz de La Palma, Spain\\
}

\date {Submitted. Accepted}
\pubyear{2001}

\begin{document}

\maketitle


\begin{abstract}
We present J band observations of a complete sample of 57 red 
galaxies selected to have \K$<$20 and R$-$\K$>$5.3. We use 
the Pozzetti and Mannucci (2000) prescriptions,
based on the R$-$K and J$-$K colours,
to separate the two dominant populations, old ellipticals and
dusty starbursts.
We find that both populations are present in the
current sample and have similar abundances,
and discuss the uncertainties in this result.  
Galactic stars comprise about 9\% of the objects.
The starburst galaxies of the present sample are found to give a contribution
to the cosmic star formation density similar to the Lyman-break galaxies.
\end{abstract}


\section{Introduction}

The Extremely Red Objects (EROs) are galaxies selected
to have very red optical-to-IR colours (such as R$-$K$>$5.3 or I$-$K$>$4). 
These extreme colours can be due either to an old stellar population
in an elliptical galaxy at redshifts between 1 and 2, or to dust 
reddening of a starburst galaxy or an AGN.
Discovered by Elston et al. in 1988, it is still not clear what is the
dominant population, and the subject is under active study.
Obtaining direct and complete spectroscopic and morphological 
information is not an easy task because EROs are
usually faint at both near-IR and optical wavelengths (K$\sim$19 and
R$\sim24$). 
As a consequence,
good spectra are available for just a few objects, and no complete
sample of EROs has good spectroscopic coverage to date. 
The limited set of available spectra (e.g.,
Soifer et al., 1999; Dey et al., 1999;
Liu et al., 2000) suggests that both type of objects are present in 
the ERO population, but the
relative abundances remain unknown.

Investigations on the morphology (Moriondo et al., 2000; 
Treu et al., 1999) and photometric redshifts (Cimatti et al., 1999) 
of composite samples have produced some evidence that the EROs with 
K$<$19 are dominated by elliptical galaxies.  The clustering properties 
of EROs with K$<$20 (Daddi et al., 2000; McCarthy et al., 2000) 
are also interpreted as a clue to the presence of more old 
ellipticals than dusty starbursts.

Recently Pozzetti and Mannucci (2000, hereafter PM2000)
introduced a method to statistically separate the two populations by 
making use of additional observations in the J band.
Old stellar populations are expected to show a strong spectral break 
between 3000 and 5000\,\AA, while 
dust reddening of a young population would produce smoother spectra.  
For redshifts between 1 and 2 this effectively separates the two populations
in the R$-$K vs. J$-$K colour plane, with starbursts having redder J$-$K
colours. A separation line can therefore be drawn in this colour-colour 
plane by using predictions from stellar synthesis models and the small 
number of observed objects with robust classifications.  
The separation is not very large, 
about 0.3 mag, so individual objects can be
misclassified because of peculiar properties or photometric 
uncertainties.  Nevertheless this method is expected to give 
reliable statistical information on the relative numbers of ellipticals and
starbursts when applied to large enough samples. 
The colours of the dusty starburst galaxies are not very sensitive to the 
possible presence of an underlying old stellar population, therefore 
the objects with composite stellar populations are expected to 
preferentially fall in the starburst region.

Thompson, Aftreth and Soifer (2000, hereafter TAS2000) observed the field of 
a radio galaxy at a redshift of 1.47, B3 0003$+$387,
to select EROs by their R$-$\K\ colours. 
They observed an area of 44.3 sq.arcmin down to a
limiting magnitude of \K=20.3.
A significant overdensity of red objects is found
in this field, in agreement with the strong clustering detected by 
Daddi et al. (2000).  We have observed this field in the J band to 
classify the objects of this complete sample as ellipticals or starbursts. 

Observations and data
reduction are described in the following section. Object selection,
astrometry and photometry in section 3; in section 4 we present the
colour-colour diagram and the relative abundances of the two populations;
the uncertainties of the classification are
discussed in section 5. In section 6 we derive some
consequences on the cosmic star formation history.

\section{Observations and data reduction}

Observations were obtained at the 3.5m Telescopio Nazionale Galileo
(TNG) in the Canary Islands. We used the Near-Infrared Camera and 
Spectrometer (NICS), a
multi-purpose instrument (Baffa et al, 2001) based on a
Hawaii 1024$\times$1024 HgCdTe array. The central pixel scale is 
0$\arcsec$.253/pix.  We used a Js filter, similar to the standard J 
filter but with a narrower bandpass (1.16 - 1.33 $\mu$m) and a 
higher throughput.
We covered the field with many short exposures (1.5 min) for a total
integration time of about 3.2 hours.
The camera has a field of about $4\arcmin\times4\arcmin$, smaller that
the TAS2000 selection field of about $7\arcmin\times7\arcmin$. 
Therefore we used a mosaic pattern to 
observe most of the EROs in the field. The telescope was dithered over 
three overlapping fields, observed one after the other
in order to obtain a final
image with a homogeneous PSF and photometric zero point.

We used a two-step reduction procedure.
A master sky image was first obtained from a median of all the available
images, and subtracted from each frame to obtain a zero-order
reduction.  The IRAF task DAOFIND was used to
detect the 50-100 brightest objects on these subtracted frames 
to make an object mask.
The second step of the procedure makes use of running sky frames 
using the 6 adjacent exposures, excluding the pixels
contained in the object mask.  
Running sky frames were subtracted from each image, and the results 
divided by a flat field obtained at sunset.

Precise dithering offsets were measured by selecting a set of about 50 
bright compact objects in the first frame and 
automatically identifying them in the following frames, accounting for 
the telescope offsets and field distortion.  The differences between 
the recovered positions of these objects in the various images give 
the offsets, with a typical error of about $0.1\arcsec$, 
less than one-half of a pixel.

The distortions from the optics were measured using a crowded stellar 
field (Hunt and Licandro, priv. comm.), reaching about 3\% at the 
corners of the array.  The distortion parameters and the measured 
offsets were used to drizzle (Fruchter \& Hook, 1997) each image 
onto its final position, creating an undistorted and co-aligned frame 
for each input image.  Given that the PSF is adequately sampled, the 
results are not very sensitive to the drop size, chosen to be 0.8.
The resulting images were combined using a clipped average, which is 
very efficient at removing cosmic ray hits and residual bad pixels.

The final PSF is about $1.1\arcsec$ FWHM. 
An accurate morphological study of the objects far from the field centre 
is prevented by the PSF degradation due to the aberration of the optics 
near the edges of the field of view.

The photometric zero point was measured by a set of standard stars from 
the ARNICA list (Hunt et al., 1998) observed at about the same airmass
as the ERO field.
We compared the observed R$-$\K\ and J$-$\K\ colours of the stars in the field
with the expected colours to check for any error in the zero points.  We
considered the objects classified as stars by Sextractor (Bertin \&
Arnouts, 1996;
i.e., with a stellarity above 0.9), and compared their colours with those
derived by integrating the stellar spectra in the Pickles (1998) library 
using the same set of filters.  A small but systematic difference between 
the observed and synthetic J$-$\K\ colour of about $-0.05$ mag was found, and 
the J-band zero point was corrected accordingly.

The total integration time in each pixel of the final image depends
strongly on the pixel position, going from an average value of
about 1 hour up to 3 hours in the small region covered by almost all 
of the images. The detection limit varies accordingly with the integration 
time.  We estimated this limit from the sky noise, after taking into account
the correlation between adjacent pixels introduced by the distortion
correction. We followed the procedure outlined in Williams et al. (1996), 
and measured the ratio between the noise in the final image and that in 
the case of no correlation. 
The latter was measured on an image constructed without the distortion 
correction. The correction factor is 1.4, and we used this in eq. 2 of 
Pozzetti et al. (1998) to estimate the real photometric errors for each 
object.  The measured 5$\sigma$ detection limit inside an aperture of 
twice the seeing, 2.2 arcsec, at the positions of the 
selected EROs (see below) varies between 21.2 and 22.5, with an
average of 22.1.

\begin{figure*}
\centerline{\psfig{figure=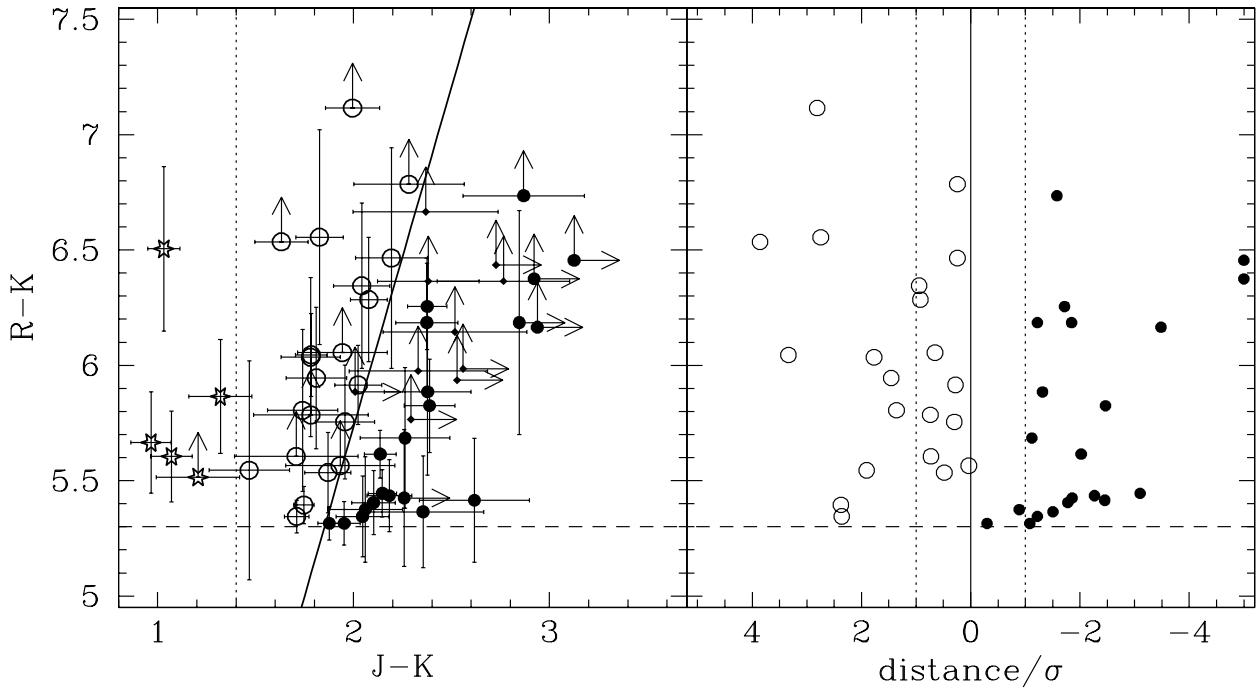,angle=-90,width=18cm}}
\caption{{\em Left panel}: R$-$\K\ vs. J$-$\K\ colour diagram of the 
selected EROs.
The thick line is the separation between ellipticals and starbursts derived
following PM2000. The horizontal dashed line is the colour threshold for
selection, the dotted vertical line the J$-$\K\ colour of the separation
between  stars and galaxies.  Open circles represent objects classified 
as ellipticals, while filled circles represent the starbursts.  Stars 
are used for the stars. 
Objects with no classification are shown as small diamonds.
{\em Right panel}: significance of the classification as elliptical or
starburst. The J$-$\K\ colour of the left panel is replaced with the
ratio between the distance of each point from the separation line and
the radius of the error ellipse in the direction perpendicular to
the line. Objects with a ratio below -5 were plotted at this value.
The dotted lines show the 1$\sigma$ limit of the classification.
}
\label{fig:rjk}
\end{figure*}

\section{Object selection, astrometry and photometry}

An object catalog down to \K$\sim$20 was made by TAS2000 and the detected 
objects measured in the R image.  The catalog contains magnitudes 
inside an aperture of 4 arcsec, much larger than the seeing and the 
intrinsic dimensions of most of the objects of interest in this paper.
We corrected the TAS2000 magnitudes for Galactic extinction using the 
Schlegel et al. (1998) value and the Cardelli et al. (1989) extinction curve.  
The resulting corrections are 0.215, 0.073 and 0.030 mag in R, J and \K, 
respectively. Using the (lower) Burstein \& Heiles
(1982) value for the Galactic extinction would result in R$-$\K\ and J$-$\K\ 
colours only 0.02 and 0.01 mag redder. 

We selected all the objects having R$-$\K$>$5.3, which corresponds to 
the expected colours of an elliptical at z$\sim$1 
(see, for example, Cimatti et al., 2000).  
There are 59 such objects, with \K\ magnitudes down to 19.9.

A common coordinate system was established among all the images
using the brightest objects, with a resulting r.m.s. 
positional uncertainty 0.2 arcsec, less than one pixel. 
This solution was used to derive the
expected positions of the EROs in the J band image.
57 of the 59 selected objects are inside the
final J band image.  The IRAF routine
PHOT was used to measure the object flux at these
positions inside a circular aperture of 4 arcsec, as in TAS2000. 
The PSF of all the
images are similar, between 0.8 and 1.1 arcsec, and the aperture large
enough that the errors on the colours due to differential aperture losses 
is below 0.05 mag.

\section{The colour-colour diagram}

We computed the expected R$-$\K\ and J$-$\K\ colours of elliptical galaxies and
dusty starbursts as in PM2000 for the present filter set: 
Harris R (0.56 - 0.74 $\mu$m), J$_s$ (1.16 - 1.33 $\mu$m) and \K\
(1.94 - 2.29 $\mu$m).  The resulting separation line can be described by:
$(\rm{J}-\rm{K}^{\prime}) = 0.34\cdot(\rm{R}-\rm{K}^{\prime})+0.05$.
This line is 0.14 mag bluer than in PM2000 mainly because the \K\
filter is centred at bluer wavelengths than a standard K filter.  
Figure~\ref{fig:rjk} shows the position of the objects around this line.
 
Objects detected in all three bands, and with J$-$\K\ colours redder than the
line were classified as starbursts.  Bluer objects were classified as 
ellipticals.
A 3$\sigma$ limiting magnitude was assigned to objects undetected 
in one band, i.e., with a measured flux below the 3$\sigma$ uncertainty 
inside the 4 arcsec aperture. 
Objects below the separation line but undetected in R, or to the left of
the line but undetected in J, were not classified.
Objects undetected in R were classified as
starbursts if their J$-$\K\ colours were larger than 2.8, corresponding
on the separation line to a R$-$\K\ colour of 8. 
Objects with J$-$\K$<$1.4 have
colours more typical of M or L dwarf stars (see, for example, Burrows
et al., 2001).  We therefore classified 5 objects having such 
colours as stars.  The morphology can be reliably studied in four of them
(see below), and they are all point sources, confirming this classification.
For the final sample of 57 red objects, 21 (37\% of the sample)
are classified as ellipticals, 21 (37\%) as starbursts, 5 (9\%) as stars. 
Ten (17\%) remain unclassified. 
Note that if the two galaxy populations have, on average, different morphology, 
the relative abundances in any magnitude-limited sample might
be affected by different selection effects, especially when the object 
detection is based on HST data.

Several objects fall near the separation line so that their
classification can be affected by the uncertainties on the colours.
The significance of the classification was estimated by the ratio of the
distance of each point from the classification line divided by the
radius of the error ellipse in the direction perpendicular to the line.
These results are shown in the right panel of Figure~\ref{fig:rjk}. 
Eleven ellipticals and 2 starbursts have a distance from the line
below 1 sigma, therefore their classification should be considered
weak.

Figure~\ref{fig:rjk} can also be used to investigate the presence 
of any correlation between the classification and the R$-$\K\ colour, as
proposed by Moriondo et al. (2000) from the morphology. In that case,
the most irregular objects tend to show the reddest colours. 
No such effect is seen in the present sample, but it should be noted 
that there is only one object with R$-$\K$>$7.

\section{Uncertainties and checks}

The PM2000 classification method suffers from some uncertainties that
should be accurately discussed and tested.

Many objects in Figure~\ref{fig:rjk} fall near the separation line, at a
distance lower than the uncertainties in their colours.
Therefore they could easily move from one side of the line
to the other because of the measurement errors.  While this is a 
problem for single objects, photometric uncertainties 
are unlikely to effect the statistical result
very much because the density of points on the two sides of the
line is similar. 

Important systematic effect could be introduced by the
uncertainty on the position of the separation line 
and on the photometric zero point.
From the separation of the regions covered by the models (see PM2000)
we estimate the former uncertainty to be about 0.1 mag in the J$-$\K\ colour,
while the latter is about 0.05 mag.
The combined effect of these on the classification can be estimated
by moving the line toward bluer or redder J$-$\K\ colours by 0.12 mag:
the fraction of ellipticals over starbursts, being 1.0 for the central
position of the line, varies from 0.65 and 1.25. 
In all the cases, the two populations are present at similar levels, 
within a factor of 35\%.

The PM2000 classification methods works for redshifts between 1 and 2.
Ellipticals at z$>$2 are expected to show red J$-$K colours
because the 4000\AA\ break enters into the J band. As a consequence,
some of the objects classified as starburst
could be ellipticals at redshifts larger than 2.
We cannot exclude this for all of the objects, but this is unlikely
to be the case for a significant number of starbursts:
not many ellipticals at z$>$2 are expected in a survey this size 
to K$\sim$20 (see, for example, Cowie et al., 1996). 

\begin{figure}
\centerline{\psfig{figure=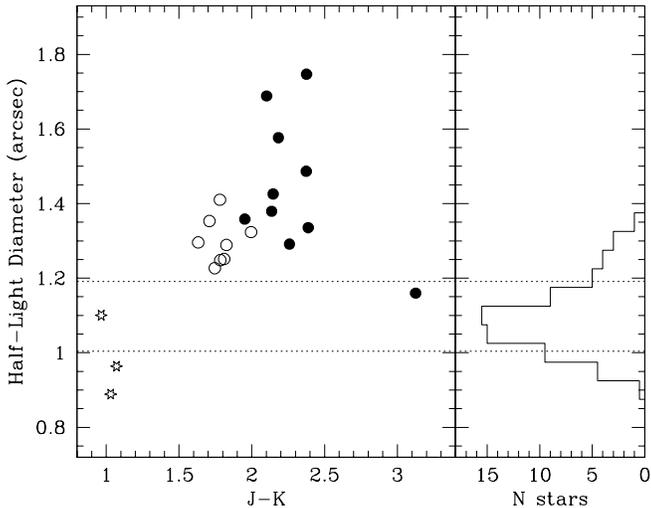,width=9cm}}
\caption{
The left panel shows the half-light diameter of the EROs vs. the J$-$\K\
colour.  Ellipticals are indicated by open circles, starbursts by solid 
circles, and stars by open stars.
The most extended objects are all
classified as starbursts based on their colours. The objects 
with J$-$\K$\sim$1 also have compact morphologies. For comparison, 
the right panel shows 
the histogram of the half-light diameter for the stars of similar magnitude.
The dotted lines
are the $\pm$1 sigma limits for the stars.
}
\label{fig:radcol}
\end{figure}

At redshifts below 1, 
the 4000\AA\ break of the elliptical galaxies is inside the R
filter bandpass.  The galaxies then have bluer R$-$\K\ colours and tend 
to be excluded from the colour-selected sample.
Nevertheless, old galaxies at z$<$1
can be selected if they have metallicities larger than
solar, making the colours redder, or if some dust is present. By making
use of the Bruzual and Charlot models (1993, version 2000) it is possible 
to see that ellipticals with metallicities 2.5 times solar at z$\sim$0.8 could
populate the  colour-colour diagram in the
lower-left part of the starburst region 
(R$-$\K$\sim$5.5 and J$-$\K$\sim$ 2) and be misclassified. It is not possible
to exclude that this is the case for some of the objects in this part of
the diagram, and this is one of the largest uncertainties for this method. 
Nevertheless the morphological information (see below) implies that 
this is not a dominant contribution: three of the objects
in this region of the diagram are among the most extended.

The morphology of the selected objects can be used as a check of the
classification method. 
The galaxies with compact morphologies can be both ellipticals and
starbursts, but we expect that the most extended or disturbed objects 
will have the colours of a dusty starburst galaxy.

The spatial extent of the objects was measured on the TAS2000 \K\ 
image, more appropriate than our J band image
because it has lower aberrations.
We have extracted the value of the half-light radius from the
Sextractor catalog both for the EROs and for the stars.
All the objects with formal error on the \K\ magnitude larger than
0.10 mag (signal-to-noise ratio lower than about 10) or with weak
classification (distance from the separation line in
Figure~\ref{fig:rjk} less than 1 sigma) were excluded.
The results for the remaining 23 objects
are shown in Figure~\ref{fig:radcol}, where the half-light diameter 
of each objects is plotted versus the J$-$\K\ colour.
As expected, the most extended objects have starburst colours. 
Some of the ellipticals with weak classifications (and therefore not 
shown in Figure~\ref{fig:radcol})
show extended morphologies: these objects could be misclassified
starbursts or objects with composite stellar populations.
It can also be seen that all the objects in Figure~\ref{fig:radcol}
with J$-$\K$\sim$1 also have compact morphologies, a strong indication that 
they are M or L dwarf stars. 

\section{The cosmic star formation history}

\begin{figure}
\centerline{\psfig{figure=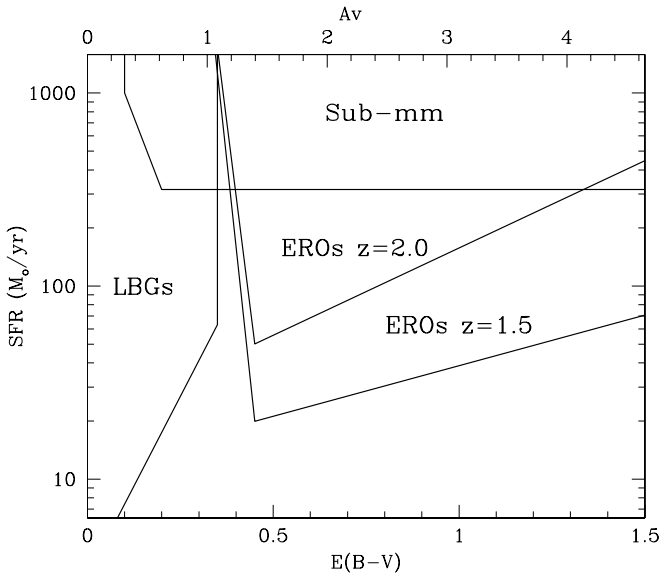,width=9cm}}
\caption{Regions of the SFR - extinction plane that can be selected by
the various search techniques. Sub-mm observations can currently detect
galaxies with any extinction but with very high SFRs. Lyman-break
galaxies (LBG) must have E(B$-$V) below about 0.3, but can have low SFRs. 
The ERO
technique covers a region of this plane with E(B$-$V) between 0.5 and 1 and
a star formation rate down to about 20 M$_{\odot}$/yr.  The exact 
values of the edges depend on many assumptions, such as the choice of 
telescope and instrument, the integration time, dust temperature, and 
redshift.  Therefore the limits should be regarded only as
indicative. Only methods aimed to detect the continuum are shown; 
techniques targeting emission lines are not included in the plot.  
}
\label{fig:selbw}
\end{figure}

The number of the starburst galaxies selected by the survey and their
individual SFRs could be used to derive the contribution to the cosmic
star formation history from this class of galaxies. This is potentially
very important as the current estimates are based on search techniques
that would miss this population. 
Many techniques to select galaxies at z$>$2 are based on the rest-frame
UV or blue continuum, and therefore tend to exclude dusty galaxies.
Most of the objects were selected by detecting a blue continuum and 
the Lyman break (e.g., Steidel et al, 1999). 
As illustrated in Figure~\ref{fig:selbw},
any galaxy with E(B$-$V) larger than about 0.3 would escape the 
selection criterion because of the lack of a very blue continuum. 
Very active, dusty galaxies can be selected by SCUBA at sub-mm wavelengths 
(e.g., Hughes, 1998).
In this case there is a lower limit rather than an upper limit
in the extinction, but the minimum SFR that can be detected at redshifts
between 1 and 2 are around a few hundred solar masses per year (see, for
example, Ivison et al., 2000).  These 
techniques are not sensitive to moderately active 
\hbox{(20$<$SFR$<$100 M$_{\odot}$/yr)}, 
moderately dusty \hbox{(0.3$<$E(B$-$V)$<$0.8)} galaxies, the
population that can be selected as EROs.

To derive a reliable value for the star formation density
is beyond the aim of this paper. 
Many large uncertainties prevent us from this: the redshift
distribution of the objects is not known; 
it is difficult to estimate
the exact cosmic volume sampled by the broad-band selection; 
the SFRs cannot be
precisely derived from the limited set of available colours;
even if observations at many wavelength were present, the value of
the SFR is usually not well constrained as it is very sensitive to the
unknown star-formation history, IMF, extinction law and amount of dust.

With these problems in mind, we derive an
order-of-magnitude estimate of the star formation density.
We assume that all the starbursts are at redshifts between 1 and 1.5.
The first limit is due to the colour selection, while the second comes 
from the luminosity limit.  The RJK colours can be reproduced by models 
of starburst galaxies in this redshift range with SFRs between 20 and 
100 M$_{\odot}$/yr and E(B$-$V) between 0.5 and 1.0. Adding the SFRs of 
the 21 galaxies in the present sample and dividing the result for the 
sampled volume of about 53000 Mpc$^3$ (H$_0$=50 Km/sec/Mpc,
$\Omega_m=1$, $\Omega_{\Lambda}=0$), 
we obtain a cosmic star formation density of about 
0.03 M$_{\odot}$/yr/Mpc$^3$; a similar value is obtained 
with H$_0$=70, $\Omega_m=0.3$ and $\Omega_{\Lambda}=0.7$.  
This value is very similar to that observed 
at higher redshift in the Lyman Break galaxies (e.g., Steidel et al., 1999). 
The uncertainties are too large to place
any reliable constraints on the galaxy formation models, and the
contribution from all the fainter undetected galaxies is not known.
Nevertheless this is an
indication that the ERO technique is not only a good method to detect
high-redshift ellipticals, but also a promising way to select a
population of starburst galaxies containing a significant fraction
of star formation and having intermediate properties between
the Lyman-break galaxies and the SCUBA objects.\\

We are grateful to L. Hunt and J. Licandro for providing the
NICS distortion parameters, to the TNG staff for support during the
observations and to L. Testi for discussion about brown dwarfs.

The object catalog can be obtained in electronic form from the web site:
www.arcetri.astro.it/$\sim$filippo/RJK

\end{document}